\documentstyle[12pt,epsf,cite]{article} 
\newcommand{\ba}{\begin{array}}
\newcommand{\ea}{\end{array}}
\newcommand{\bd}{\begin{displaymath}}
\newcommand{\ed}{\end{displaymath}}
\newcommand{\be}{\begin{equation}}
\newcommand{\ee}{\end{equation}}
\newcommand{\bea}{\begin{eqnarray}}
\newcommand{\eea}{\end{eqnarray}}

\def\dis{\displaystyle}
\def\barr{\begin{array}}
\def\earr{\end{array}}
                               
                              \def\gev{\: \rm GeV} 
                              \def\tev{\: \rm TeV} 
                              
                              \def\fb {\: \rm fb}
\newcommand{\beqn}{\begin{eqnarray}}
\newcommand{\eeqn}{\end{eqnarray}}

\def\ra{\rightarrow}

\def\a{\alpha}

\def\b{\beta}

\def\q2 {q^2}

\def\r {\rightarrow}

\def\er {\tilde{e}_R}

\def\el {\tilde{e}_L}
\def\N10{\widetilde \chi_1^0}
\def\Cp1{\widetilde \chi_1^+}
\def\Cm1{\widetilde \chi_1^-}
\def\C1pm{\widetilde \chi_1^\pm}
\def\Ntwo{\widetilde \chi_2^0}

\def\sneu{\tilde \nu}

\def\mpT{p_T \hspace{-1em}/\;\:}

\def\lsim{\:\raisebox{-0.5ex}{$\stackrel{\textstyle<}{\sim}$}\:}
\def\gsim{\:\raisebox{-0.5ex}{$\stackrel{\textstyle>}{\sim}$}\:}
\begin{document}
\begin{flushright}
MRI-P-020802\\[2ex]
{\large \tt hep-ph/0208240}

\end{flushright}

\begin{center}
{\Large\bf Signals of anomaly mediated supersymmetry breaking in an
$e^- \gamma$ collider}\\[10mm]
{\bf Debajyoti Choudhury$^{a,}$\footnote{debchou@mri.ernet.in}, 
Dilip Kumar Ghosh$^{b,}$\footnote{Dilip.Ghosh@cern.ch}, {\rm and}
Sourov Roy$^{c,}$\footnote{roy@physics.technion.ac.il}}\\[4mm]
{\em $^a$Harish-Chandra Research Institute, Chhatnag Road, Jhusi \\
Allahabad - 211 019, India} \\[5mm]
{\em $^b$National Center for Theoretical Sciences, National Tsing Hua University \\
Hsinchu, Taiwan , R. O. C.}\\[5mm]
{\em $^c$Department of Physics, Technion - Israel Institute of Technology \\
Haifa 32000, Israel} \\[10mm]
\end{center}
\begin{abstract}
We study the signatures of minimal anomaly mediated supersymmetry breaking 
in an $e^- \gamma$ collider. We demonstrate that the associated production 
of a sneutrino with the lightest chargino leads to a substantially large 
signal size. The background is negligibly small, though. Even more 
interestingly, a measurement of the fundamental supersymmetry 
breaking parameters could be possible. 
\end{abstract}

\vskip 1 true cm

\noindent
PACS NOS. : 12.60.Jv, 13.10.+q, 14.80.Ly

\setcounter{footnote}{0}

\newpage

\section{Introduction}
          \label{sect:introd}
The question of supersymmetry (SUSY) breaking is of burning relevance in 
high energy physics today. The most general version of the minimal 
supersymmetric standard model (MSSM), with the attendant
large number of arbitrary SUSY
breaking soft parameters, makes itself untractable for experimental 
searches. However, if the mechanism of SUSY breaking were known,
then it would be 
possible to reduce these large number of parameters into a much smaller set.
With the corresponding ordering in the mass spectrum, 
the nature of the lightest supersymmetric particle (LSP) would be 
determined, and the decay chains established, thereby
making the model much more predictive.
Several such supersymmetry breaking mechanisms 
have been discussed in the literature, alongwith their
distinctive phenomenological signatures.
Anomaly mediated supersymmetry breaking (AMSB) 
\cite{randall-sundrum,giudice-luty} is one such possibility and 
has attracted a lot of attention in recent times. 
Building on the basic idea, wherein 
the supersymmetry breaking is conveyed to the observable sector 
by the super-Weyl anomaly, a whole class of models have 
been constructed~\cite{bagger,pomarol,kribs,katz,jack,carena,chacko,dedes,
arkani-hamed,chacko1,jackjones,kaplan,weiner,campos,wild,anisimov,
luty-sundrum,harnik},
and many of the phenomenological implications 
discussed~\cite{kribs,campos,rat_stru_pom,feng,wells,fengmoroi,ssu,paige,
tata,konar,barr,dkgprsr,ghosh,gamma_gamma}.
For example,
the characteristic signatures of the minimal anomaly mediated supersymmetry 
breaking model (mAMSB) have been studied in the context of hadronic 
colliders~\cite{feng,wells,fengmoroi,ssu,paige,tata,konar,barr},
as well as for high energy linear colliders, whether of the 
$e^+e^-$ type~\cite{dkgprsr,ghosh}, or photon-photon 
colliders~\cite{gamma_gamma}. In this paper we study 
the unique signatures of the mAMSB model in an $e^-\gamma$ collider.

In its original version, the AMSB scenario consisted of a higher-dimensional
supergravity theory  wherein the hidden sector and observable sector 
superfields are localized on two distinct parallel three-branes 
separated by a distance  $\sim r_c$ ($r_c$ is the compactification radius) 
in the extra dimension \cite{randall-sundrum}. 
Below the compactification scale 
($\mu_c \sim r_c^{-1}$),  only four-dimensional supergravity fields 
are assumed to propagate in the bulk. Even if there are additional bulk fields 
above this scale, their effects would, typically, be 
suppressed by a factor $e^{-m/{\mu_c}}$.
As there is no tree level coupling between the hidden sector fields and 
those in the observable sector, it had been assumed that the  
flavor changing neutral current (FCNC) processes would be suppressed naturally,
thereby solving a longstanding problem of supersymmetric theories. 
Recently, though, 
it has been shown that the physical separation between the visible and hidden 
sectors in extra dimension is not sufficient to suppress the FCNC processes
except in some special cases \cite{anisimov}. However, it is 
possible to construct models of AMSB, even  
in a completely four-dimensional framework, 
that can circumvent the flavour problem \cite{luty-sundrum,harnik}. 
For the purposes of the present study, 
we do not need to delve into such details and would rather be
concentrating on the minimal scenario of Ref. \cite{randall-sundrum}.

If one were to describe the AMSB scenario in terms of a 
four-dimensional effective theory, applicable below $\mu_c$, 
then a rescaling transformation can be defined so as to eliminate, 
from the classical Lagrangian,
any tree-level interaction (except for the $\mu$-term in the superpotential) 
connecting the supergravity fields with the visible sector matter fields.
Such a scaling transformation, however, is anomalous and hence 
supersymmetry breaking is communicated from the hidden sector to the visible 
sector through the super-conformal anomaly \cite{randall-sundrum}. The 
supersymmetry breaking soft mass parameters for the gauginos and the scalars 
are generated at the same order in the corresponding gauge coupling strength. 
The analytical expressions for the scalar and gaugino masses,
in terms of the supersymmetry breaking parameters, are renormalization group 
(RG) invariant, and, thus, can be computed at the low-energy scale 
in terms of the appropriate beta functions and anomalous dimensions. 
The minimal AMSB scenario suffers from a 
glaring problem though. At low energies, it is beset with 
the existence of tachyonic sleptons. Several solutions to this problem
exist. In this work, we shall consider the 
minimal AMSB model wherein a constant term $m^2_0$ is added to all the scalar 
squared masses thereby making the slepton mass-squareds sufficiently positive.
While this may seem to be an {\em ad hoc}\ step, 
models have been constructed that naturally lead to such an eventuality.
A consequence is that the RG 
invariance of the expression for scalar masses is lost and hence 
one needs to consider the corresponding 
evolution down to the electroweak scale.

The minimal AMSB model has several unique features. The gravitino is 
very heavy, its mass being in the range of tens of TeV. Left and right 
selectrons and smuons are nearly mass-degenerate while the staus split into 
two distinct mass eigenstates. But perhaps the most striking feature is 
that both the lightest supersymmetric particle (LSP) $\N10$ and 
the lighter chargino ($\C1pm$) are predominantly Winos and hence  
nearly mass degenerate. Loop corrections as well a small 
gaugino-Higgsino mixing at the tree level do split the two, but the 
consequent mass difference is very small: $\Delta M < 1 {\rm GeV}$.
The dominant decay mode of the lighter chargino is $\C1pm \r \N10 +
\pi^\pm$ and this long-lived chargino would typically
result in  a heavily ionizing charged
track and/or a characteristic soft pion in the detector \cite{gunion}. 

Signals of supersymmetry in an $e^- \gamma$ collider have been discussed 
in various contexts~\cite{robinett,cuypers,peterson,kon,choudhury,kiers,
barger-han,ghosal,ghosh-ray}. In this paper, we consider the process 
$e^- \gamma \r \sneu \Cm1$ to look for signals of anomaly mediated 
supersymmetry breaking in such a  collider. 
This choice has some advantages over the possibilities at an $e^+ e^-$ 
collider. The dominant production channel at the latter machine, namely
a $\tilde \chi_1^+ \tilde \chi_1^-$ pair, 
is notoriously difficult to tag onto, and 
even if detected, is hardly amenable to mass determination. And as the 
sleptons tend to be significantly heavier than $\tilde \chi_1^+$, at least 
for a very large part of the parameter space, the $\sneu \Cm1$ threshold
tends to be quite a bit lower than those for $\tilde \ell^+ \tilde \ell^-$
or $\sneu \sneu^*$. Moreover, the cross section for the first-mentioned 
process is, generically, much higher,  and even more so close to the 
production threshold.

Once produced, the sneutrino may decay into either an ($e^- \tilde \chi_1^+$)
pair or a ($\nu \tilde \chi_1^0$) pair. Concentrating upon the former, 
we are left with a fast $e^-$ (which serves as the trigger), 
two heavily ionizing charged tracks 
coming from the long-lived $\C1pm$ and/or two visible soft pions with 
opposite charges\footnote{While it might not very easy to measure the 
pion charges, they are nonetheless distinguishable, if only in a statistical 
sense, from their rapidity distribution.}
and a large missing transverse momentum ($\mpT$). 
This is a very unique and distinct 
signature of anomaly mediated supersymmetry breaking and does not 
readily arise in either of mSUGRA or GMSB scenarios. 

What of other channels at an $e^- \gamma$ collider itself? Continuing 
with the same production process, had the selectron decayed into 
the alternate channel, namely a ($\nu \tilde \chi_1^0$) pair, we would 
have been left with $\pi^- + \mpT$ final state, possibly associated 
with a single heavily ionized track. The lack of a reliable trigger 
renders this channel of little use. 
The associated production of a 
left\footnote{Note that $e^- \gamma \r \er \N10$ is suppressed in this model
	on account of  $\N10$ having a  vanishingly small Bino component.}
selectron and the lightest neutralino
($e^- \gamma \r \el \N10$), on the other 
hand, would result in the same signal ($e^- + \mpT$)
as in the case of mSUGRA. Though eminently detectable, this 
is of hardly any use in establishing the nature of 
supersymmetry breaking, and hence we desist from discussing it any further.
It is, of course, true that, for some choices of the SUSY
parameters, the selectron may also decay into the heavier neutralinos
($\el \r e^- \Ntwo$) thereby leading to final 
states with multifermions and $\mpT$. 
However, the signal cross section in 
this case will be small on account of the suppressions due to the various 
branching ratios. And finally, one may also consider 
associated production of heavier charginos (with sneutrino) 
and neutralinos (with selectron) and their subsequent cascade decays.
The signals for such processes would be rather 
complex, though. And the production cross sections typically smaller, 
both on account of reduced phase space as well as smaller couplings.
It may, thus, be said that the channel of our choice is 
the simplest as well as the most promising one.

The paper is organized as follows. In Sec.2, we give a brief description
of the $e^- \gamma$ collider.
Sec.3 discusses the spectrum and the couplings within the mAMSB model and 
the constraints on the parameter space from various experimental and 
theoretical considerations. Numerical results of our computations and
their discussions are presented in Sec.4. In Sec.5, we discuss the 
possibility of determining the supersymmetry breaking parameters.
Finally, we conclude in Sec.6.

\section{$e^- \gamma$ collider and the photon spectrum}

While it is quite apparent that maximizing 
the signal cross sections implies using perfectly polarized electron 
and photon beams, in reality, though, perfect polarizations is almost 
impossible. Furthermore, even near monochromaticity 
for high energy photon beams is
extremely unlikely. In fact, the only known way to obtain 
very high energy photon beams is to induce laser back-scattering 
off an energetic $e^\pm$ beam~\cite{telnov}. The reflected 
photon beam carries off only a fraction ($y$) of the $e^\pm$ 
energy with 
\be
\barr{rcl}
y_{\rm max} & = & \dis \frac{z}{1 + z} 
	\\[2ex]
z & \equiv & \dis 
	\frac{4 E_b E_L}{m_e^2} 
                 \cos^2 \frac{\theta_{b L}}{2} \ ,
\earr
\ee
where $E_{b (L)}$ are the energies of the incident electron(positron) beam 
and the laser respectively and $\theta_{b L}$ is the incidence angle.
In principle, one can increase the photon energy by increasing the 
energy of the laser beam. However, a large $E_L$ (or, equivalently, a
large $z$) also enhances the probability of electron positron
pair creation through laser and scattered-photon interactions, and 
consequently results in beam degradation. 
An optimal choice is $z = 2(1 + \sqrt{2})$,
and this is the value that we adopt in our analysis.

The cross-sections for a realistic electron-photon collider 
can then be obtained by convoluting the fixed-energy 
cross-sections $\hat{\sigma}({\hat s}, P_{\gamma}, P_{e^-})$
 with the appropriate photon spectrum: 
\be
\sigma(s) 
= \int {\rm d} y \; {\rm d} \hat s \; \:
    \frac{{\rm d} n}{{\rm d} y} (P_{b}, P_L) \; \:
\hat{\sigma}({\hat s}, P_{\gamma}, P_{e^-}) \; \delta(\hat s - y s) \ ,
	\label{csec_convolution}
\ee
where the photon polarization is itself a function of $P_{b, L}$ and the 
momentum fraction, viz. $P_{\gamma} = P_{\gamma}(y, P_b, P_L)$. For simplicity,
we shall only consider circularly polarized lasers scattering off 
polarized electron (positron) beams. The corresponding
number-density $n(y)$ and average 
helicity for the scattered photons are then given by~\cite{telnov}
\be
\barr{rcl}
\dis \frac{dn}{dy} &=&  \dis 
	\frac{2 \pi \alpha^2}{m_{e}^2 z \sigma_C} C(y) 
   \\[2ex]
P_\gamma (y) &=& \dis 
	\frac{1}{C(y)} \bigg[ P_b \bigg\{ \frac{y}{1-y} + y(2 r -1)^2
\bigg \} - P_L (2 r -1) \bigg( 1 - y + \frac{1}{1-y} \bigg) \bigg] 
    \\[2ex]
C(y) &\equiv& \dis 
	\frac{y}{1-y} + (1 -y) - 4r(1-r) - 2P_b P_L rz (2r -1)(2 -y) \ ,
\earr
	\label{photon_spectrum}
\ee
where $r \equiv y / z / (1 - y)$ and the total 
Compton cross-section $\sigma_C$ provides the normalization.

A further experimental issue needs to be concerned at this stage. While
the photon spectrum of eqn.(\ref{photon_spectrum}) has a long low-energy 
tail, in a realistic situation it might be that they cannot participate 
in any interaction.
There is a one to one relationship
between the energy of the back-scattered photons
and their angle with respect to the direction of the initial electron:
harder photons are emitted at smaller angles
whereas softer photons are emitted at larger angles.
For example, for small deflection angles, we have
\be
\theta_\gamma (y) \simeq {m_e \over E_b} \sqrt{{z \over y}-z-1} \ .
\ee
Since the photons are distributed according to an effective 
spectrum (eqn.\ref{photon_spectrum}),
this relation effectively throws  out the low energy photons,
these being produced at too wide an angle to contribute significantly
to any interaction. The exact profile of this effective spectrum, though,
is not so simple and depends somewhat on the shape of the electron beam,
and the conversion distance,
{\em i.e.} the distance between the interaction point
and the point where the laser photons are back-scattered.
In the absence of a detailed (and machine specific)
study of this effect, we are, unfortunately, not in a position 
to include it in our simulations. However, as the study of 
Ref.~\cite{choudhury} had indicated, and as we have checked, neglecting 
this effect does not change the total signal cross section to any significant
extent. Elimination of the low energy photons, however, can help in 
reducing the backgrounds, and thus our approximation
is a {\em conservative} one.

Before we end this section, we would like to point out that while 
perfect polarization for the laser beam is relatively easy to obtain, the 
same is not true for electrons or positrons. In our studies with 
polarized beams, we shall then
use $|P_L| = 1, |P_{b}| = |P_{e^-}| = 0.8$ which, once again, reflects a 
conservative choice. Since we seek to produce the sneutrino, it follows 
that the $e^-$ should be preferentially left-polarized, or $P_{e^-} = -0.8$.
Similarly, choosing the laser and the $e^\pm$ beam to be oppositely 
polarized ($P_L \times P_{b} < 0 $) improves
the monochromaticity of the outgoing photons~\cite{klasen}. For the sake 
of completeness, we shall use both choices of polarizations consistent 
with $P_L \times P_{b} < 0 $. While the total cross sections are 
obviously dependent on the polarization choice, the efficiencis of the 
kinematical cuts are expected not to be.

\section{Model parameters and constraints}

The minimal AMSB model has a high degree of predictivity as 
it is described by just three parameters (apart from the SM parameters,
of course): the gravitino mass $m_{3/2}$, the common scalar mass
parameter $m_0$ and $\tan\b$, the ratio of the two Higgs vacuum
expectation 
values. In  addition, there is a discrete variable, namely 
the sign of the Higgs mass term ($\mu$). As mentioned earlier, 
the soft supersymmetry breaking terms in the effective Lagrangian are then 
determined solely in terms of the gauge ($g_i$) and Yukawa ($y_a$)
couplings.
Denoting the generic beta-functions and anomalous dimension
by $\b_g(g,y) \equiv dg/dt$, 
$\b_y(g,y) \equiv dy/dt$ and $\gamma(g,y) \equiv dlnZ/dt$ ($t$ being 
the logarithmic scale variable) respectively, we have, 
for the gaugino ($\lambda$)
masses
\be
M_\lambda = {\b_g \over g}m_{3/2},
\ee
where the appropriate gauge coupling and $\beta$-function are to be 
considered. Similarly, for the trilinear soft breaking parameters, one has
\be
A_y = {\b_y \over y}m_{3/2} \ .
\ee
The scalar masses, on the other hand, receive contributions from more than 
one source. Apart from the individual contributions from each of the 
relevant gauge couplings, there is also the universal contribution
$m_0^2$. 
Symbolically, then,
\be
\label{scalarmass}
m^2_{\tilde f} = m^2_0 - {\frac 1 4} \sum 
{\left({{\partial \gamma} \over {\partial g}}
\b_g + {{\partial \gamma} \over {\partial y}} \b_y \right)} m^2_{3/2} 
\ee
The detailed expressions for the gaugino masses at the one loop level and
the 
squared masses for the Higgs and the other scalars at the two loop level 
can be found in Refs. \cite{wells,ghosh,utpal}.

In our analysis, we use two-loop renormalization group equations 
(RGE) \cite{martin} to evolve the gauge and Yukawa couplings from the 
unification scale ($M_G \sim 2 \times 10^{16}$) down to 
the electroweak scale. For the gauge and Yukawa couplings, 
the boundary conditions are determined at the weak scale 
(with $\a_3(M_Z) \approx 0.118$), while for the scalar masses 
these are given at the unification scale vide eqn.(\ref{scalarmass}).
The magnitude of the Higgsino mass parameter $\mu$ is computed 
from the complete one-loop effective potential \cite{effective} 
and imposing the requirement of radiative electroweak symmetry breaking. 
The optimal choice for the renormalization scale is expressible in terms 
of the the masses of the top-squarks, 
viz. $Q^2 = m_{\tilde t_1} m_{\tilde t_2}$.
We also include the supersymmetric QCD corrections
to the bottom-quark mass \cite{bottomqcd} since this plays a
significant role for large $\tan\b$. 

A particularly interesting feature of the mAMSB model is that 
the ratios of the gaugino mass parameters, at low energies, turn out to
be
\be
|M_1| : |M_2| : |M_3| \approx 2.8 : 1: 7.1 \ .
\label{eq:ratio}
\ee
In eqn(\ref{eq:ratio}),  $M_1$, $M_2$ and $M_3$ refer to the 
$U(1)$, $SU(2)$ and $SU(3)$ gaugino
mass parameters respectively. An immediate consequence is that 
the lighter chargino $\C1pm$ and the lightest neutralino
$\N10$ are both almost exclusively a Wino and, hence, 
nearly degenerate. A small mass difference is generated though 
from the tree-level 
gaugino-Higgsino mixing as well as from the one-loop corrections 
to the chargino and
the neutralino mass matrices \cite{wells}. The mass splitting
has an approximate form:

\be
\begin{array}{rcl}
\Delta M \equiv m_{\tilde \chi_1^+} - m_{\tilde \chi_1^0}
         & = & \dis 
	\frac{ M_W^4 \tan^2\theta_W}{(M_1 - M_2)\mu^2 } \sin^2 2\beta
\bigg [1+ {\cal O } \left(\frac{M_2}{\mu},\frac{M^2_W}{\mu M_1}
\right)\bigg ] 
      \\[1.2ex]
& + & \dis \frac {\alpha M_2}{\pi\sin^2\theta_W}\bigg
[f\left(\frac{M_W^2}{M_2^2} \right)- \cos^2\theta_W f\left(\frac{M_Z^2}
{M_2^2}\right)\bigg ],
\end{array}
\label{eq:delm}
\ee
with
\be
f(x) \equiv
-\frac{x}{4}+\frac{x^2}{8}\ln(x) +\frac{1}{2}\left(1+\frac{x}{2} \right)
\sqrt{4x-x^2} \; \tan^{-1}\left(\frac{ (1-x) \sqrt{4x-x^2}}
                                     { 3 x - x^2} \right) \ .
\ee
For the range of $m_0$ and $m_{3/2}$ that we would be 
considering in our analysis,  $\Delta M \lsim$ 500 MeV. 
And, for very large $M_2$, the mass difference
reaches an asymptotic value of $\approx$ 165 MeV.

To determine the parameter space allowed to the theory, 
several experimental constraints need to be considered, the most 
important ones being:
\begin{itemize}
\item $\tilde \chi_1^0$ must be the LSP;
\item  $m_{\C1pm} >$ 86 GeV, when this chargino almost degenerate with 
the lightest neutralino \cite{aleph}\footnote{The lower limit mentioned
in that paper is valid for heavier sfermion masses and is slightly above
(88 GeV) the value we consider here.}.
\end{itemize}
The last constraint serves to rule out relatively low values of $m_{3/2}$ 
{\em irrespective} of the value of $m_0$. The width of this disallowed
band depends, of course, on $\tan \beta$ and $sgn(\mu)$. The first 
constraint, on the other hand, restricts the parameter space through a linear 
relation: $m_{3/2} < a m_0 + b$, where the numerical values of the 
constants $a, b$, depend, once again, on $\tan \beta$ and $sgn(\mu)$.
Typically, the maximum 
possible value of $m_{3/2}$ for a given $m_0$ is a decreasing function of 
$\tan\b$. 
Note that the LEP2 constraints on the lighter stau mass, namely 
$m_{{\tilde \tau}_1} > 82 $ GeV \cite{delphi}, are subsumed by the ones 
listed above. 
A detailed discussion of such issues  can be found in Ref.\cite{utpal}. 

Apart from direct bounds, one must also consider the constraints
imposed on virtual exchange contributions to low energy observables.
The recent measurement of the muon anomalous magnetic
moment $(g_\mu -2)$ is a case in point. 
The resultant constraints on the mAMSB model parameters 
have been considered by several 
authors~\cite{fengmoroi,utpal,g-22.61,g-22.63}. The numerical 
results of these papers need to be modified though. For one, 
the light by light hadronic contribution to
$(g_\mu -2)$ has since been reevaluated
resulting in a reversal of the sign of this particular 
contribution~\cite{changeing-2} and a consequent reduction 
of the discrepancy with the SM result. Even more recently, the E821
experiment has published new data that confirms their earlier result 
while increasing the precision significantly \cite{preciseg-2}. 
It must be borne in mind, though, that the calculation 
of the SM contribution to $(g_\mu -2)$ is beset with many remaining
theoretical uncertainties, and, hence, any such constraint must be 
treated with due circumspection.
Constraints are also derivable~\cite{fengmoroi,g-22.61,g-22.63} 
from the measurement of the rare decay rate 
$\Gamma(B \r X_s \gamma)$, but, once again, they are not too restrictive 
and many a loophole exists. 
Additional bounds may exist if one demands that 
the electroweak vacuum corresponds to the
global minimum of the scalar potential \cite{samanta,gabrielli}. 
This restriction, however, can be evaded as long as 
it can be ensured that the local minimum has a life time longer 
than the present age of the Universe \cite{vacuum}.

\section{Signal and Backgrounds}
\begin{figure}[hbt]
\vspace*{-3.5cm} \hspace*{1cm}
\centerline{
\epsfxsize=20cm\epsfysize=30.0cm
                     \epsfbox{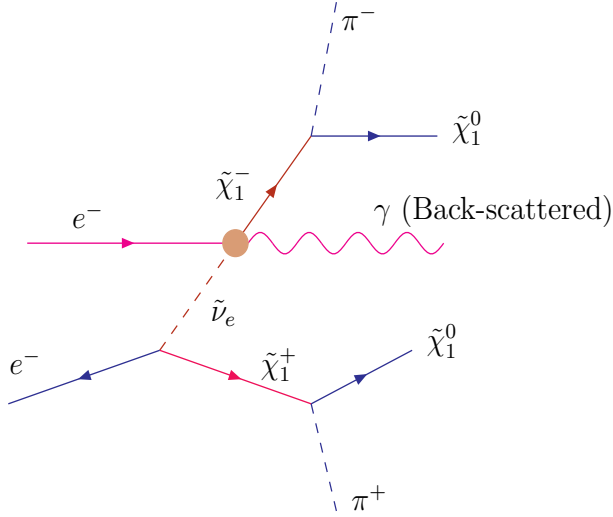}
}
\vspace*{-20.0cm}
\caption{\sf
Schematic diagram for the signal process $e^-\gamma \to \tilde\nu_e 
\tilde\chi^-_1 \to e^- +\pi^+\pi^- + \mpT $. We define $ \tilde\chi^-_{1}$
as the primary chargino and  $ \tilde\chi^+_{1}$ 
as the secondary one.}
\label{fig-feyn}
\end{figure}

As explained in Section~\ref{sect:introd}, we would be focussing on 
the production process $e^- \gamma \r \sneu \Cm1$. 
The sneutrino may subsequently decay into a ($e^- + \Cp1$)
pair or through the invisible ($\nu_e + \N10$) 
channel\footnote{The charged slepton is always  heavier than the 
	sneutrino on account of electroweak $D$-term contribution.}
with the branching fractions BR$(\tilde \nu_e \rightarrow \nu_e + \N10) 
\approx 33\%$ and BR$(e^- + \Cp1) \approx 66\%$ 
relatively well determined on account of the $\Cp1$ and the $\N10$ 
being predominantly Winos. 
Decays into the heavier charginos(neutralinos) are kinematically 
forbidden over most of the parameter space and, even if they are allowed, the 
branching ratios in those channels are very small. And while three body 
decays of the form $\sneu \r e^- \nu_\tau {\tilde \tau}^+_1$ are allowed
in principle, in practice they are too small to be of any relevance.  
For reasons of detectability, we choose to neglect the invisible 
channel and concentrate on the electron-chargino one.
The two charginos subsequently decay into a neutralino-pion pair each. 
The entire schematics of the signal is presented in Fig.~\ref{fig-feyn}.
In subsequent discussions, we shall denote $ \tilde\chi^-_{1}$
as the primary chargino and  $ \tilde\chi^+_{1}$ (arising 
from the decay of $ \sneu$) as the secondary chargino. 

It turns out that, for most of the mAMSB parameter space,
the two charged pions are well separated from each 
other. In particular, when $ m_{\tilde\chi^-_1}$ is small
compared to both $\sqrt{s}_{e\gamma }$ and $m_{\tilde\nu_e}$, the 
boosts imparted to the charginos make the pions appear almost 
back-to-back. The neutralinos, on the other hand, escape detection, 
thereby giving rise to an imbalance in momentum.

The signal, then is, 
\be
e^-\gamma \ra \tilde\nu_e \tilde\chi^-_1 \ra e^- +\pi^+\pi^- + \mpT \ ,
\ee
with the energetic electron serving as the trigger for the event. 
The relatively small decay width of the charginos is manifested 
in heavily ionizing charged tracks (one for each chargino) 
terminating inside the detector after traversing a macroscopic distance
and ending in a soft pion (with $p_T > 200$~MeV) in the Silicon Vertex
Detector (SVD) located very close to the beam pipe. 
The probability that the chargino decays before travelling a distance 
$\lambda $ is given by $P(\lambda) = 1 - exp(-\lambda/L)$, where 
$L = c\tau (\beta \gamma )$ is the average decay length of the chargino. 
Although a large fraction of the events do turn out to be associated with 
a $\tilde\chi^\pm_1$ decay that is so prompt as to make the charged track 
nearly invisible (the end product---soft $\pi$---is still detectable), a 
substantial number of events do have a reasonably large decay lengths for 
which the displaced vertex $X_D$ may be visible. Consequently, one could 
find a thick ionizing charged track in the first few layers of the SVD with 
the track terminating to give off a soft pion, 
which can be observed through its 
impact parameter. In the worst case, when the chargino track is not seen, 
our signal can still be observed by looking at the soft pion impact parameter 
$b_{\pi}$.

We select the signal events according to the following criteria:
\begin{itemize}
\item The transverse momentum of the electron must be large enough: 
	$p^e_T > 20$~GeV.
\item The transverse momentum of the pions must satisfy 
	$p^\pi_T > 0.2\gev$.
\item The total energy of the pions should not be very large 
      though: $E^\pi < 2 \gev$.
\item The electron and both the pions must be relatively central, i.e.
      their pseudorapidities must fall in the range
       $\mid \eta^{e,\pi}\mid < 2.5$.
\item The electron and the pions must be well-separated from each other:
      i.e. the isolation variable 
	$\Delta R \equiv\sqrt{(\Delta\eta)^2 + (\Delta\phi)^2}$ (where 
	$\eta$ and $\phi$ denote the separation in rapidity and the 
	azimuthal angle respectively) should satisfy 
	$\Delta R > 0.4 $ for each combination.
\item The missing transverse energy $\mpT > 20$~GeV.
\end{itemize}
Any heavily ionizing charged track would be an additional bonus. While 
the rationale behind most of the kinematical requirements listed above 
is self-evident, the importance of the upper bound on the pion 
energy would become clear shortly.

\subsection{The SM background}
Before we discuss the ensuing signal cross section, let us 
first examine the possible Standard Model backgrounds. This would 
also help put into perspective the cuts listed above. A heavily ionizing 
track may arise only from the motion of a relatively massive charged 
object. The only particles in the SM spectrum that fit the bill are ones 
that decay promptly (and hence leave no ionizing track).
Thus, if macroscopic tracks are seen , the signal is essentially 
free of backgrounds. 

What if no tracks are seen at all? The signal then comprises of 
just a hard electron and two soft pions and accompanied by a 
measure of missing energy. The backgrounds to this are manifold:
\begin{itemize}
\item \underline{$e^- \gamma \to e^- \pi^+ \pi^- \nu_i \bar \nu_i$}:
   While an accurate calculation of this is a difficult task, a fair 
   estimate may be made by the use of scalar-QED (as well as scalar QFD).
   To improve accuracy, form factors may be used to parametrize the 
   vertices involving the pions. It is here that the upper bound on the 
   pion energy, as discussed above, assumes importance. Since the pion-photon
   vertices (as well as the pion-$Z$ or pion-$W$ vertices) arise from 
   derivative couplings, a process such as the one we are describing would be 
   heavily biased towards energetic pions. With the cuts that we have 
   imposed, the total cross section of this particular background is 
    $\lsim {\cal O}(10^{-3} \fb)$ and hence entirely negligible.
\item \underline{$e^- \gamma \to e^- q \bar q \nu_i \bar \nu_i$} with 
   the quarks hadronizing into soft pions: \\
   The largest contribution to the partonic process accrues from the production
   mode $e^- \gamma \to e^- Z Z$ with a $Z$ each decaying into quarks 
   and neutrinos. It is easy to see that the quarks from such processes 
   would be very energetic and would not lead to just a pair of soft pions
   unaccompanied by any other hadronic activity. As for the continuum process 
   $e^- \gamma \to e^- q \bar q Z$ (with $Z \ra \nu_i \bar \nu_i$), it is 
   a ${\cal O}(\alpha_W^4)$ one and hence highly suppressed. \\
   It is often argued that a signal with just a pair of soft pions, an electron    trigger  and no macroscopic ionization track would suffer from large, 
   and almost incalculable, QCD backgrounds. However, as explained above, 
   the particular {\em electroweak} process that can give rise to a event 
   topology similar to our signal, is highly suppressed and even 
   non-perturbative QCD corrections cannot enhance it to significant levels.
\item \underline{$e^- \gamma \to e^- \tau^+ \tau^-$} with the taus decaying
   appropriately:\\
	While estimating this background, it is imperative that the 
	$\tau$-polarization information be retained in the intermediate 
	stages of the calculation, especially since the distribution of
	the decay products depend crucially on the polarization. We include 
	all decay channels that involve two charged pions satisfying 
	our selection criteria. Since some of the events thus included may 
	be vetoed by other criteria (such as the absence of a $\pi^0$), our 
	choice is a conservative one. On imposition of our cuts, 
	the surviving background is $\lsim 0.05 \fb$, with the exact 
	numbers depending on the polarization choice.
\item \underline{$e^- \gamma \to e^- W^+ W^-$ with $W \to \tau \to \pi$ }:\\
	While the $W$-pair production cross section is sizeable, the 
        $\tau$'s and hence the $\pi$'s tend to be harder. This serves to 
	suppress the background to below ${\cal O}(10^{-3} \fb)$.
\end{itemize}

\subsection{The signal profile}
As we have seen in the previous section, we may
safely conclude that, even in the absence of macroscopic 
ionization tracks, our selection criteria serves to remove virtually all 
of the SM background. What of the signal, then? Clearly, both the 
number of signal events and the kinematical distributions thereof would 
depend crucially on the sneutrino and chargino mass. We will illustrate 
the case for two widely separated points in the parameter space:
\begin{itemize}
\item[{\bf A.}] $m_0 = 210\gev, m_{3/2} = 37\tev, \tan\beta = 5$ and 
$\mu > 0$\\
which leads to $(m_{\Cp1}, m_{\N10}, m_{\tilde \nu}) 
		= (119.64, 119.39, 139.11) \gev$; and
\item[{\bf B.}] $m_0 = 450\gev, m_{3/2} = 55\tev, \tan\beta = 30$ and 
$\mu > 0$\\
leading to $(m_{\Cp1}, m_{\N10}, m_{\tilde \nu}) 
		= (186.42, 186.23, 390.07) \gev$.
\end{itemize}
\begin{figure}[!b]
\vspace*{20ex}
\centerline{\hspace*{-20ex}
\epsfxsize=10cm\epsfysize=11.0cm
                    \epsfbox{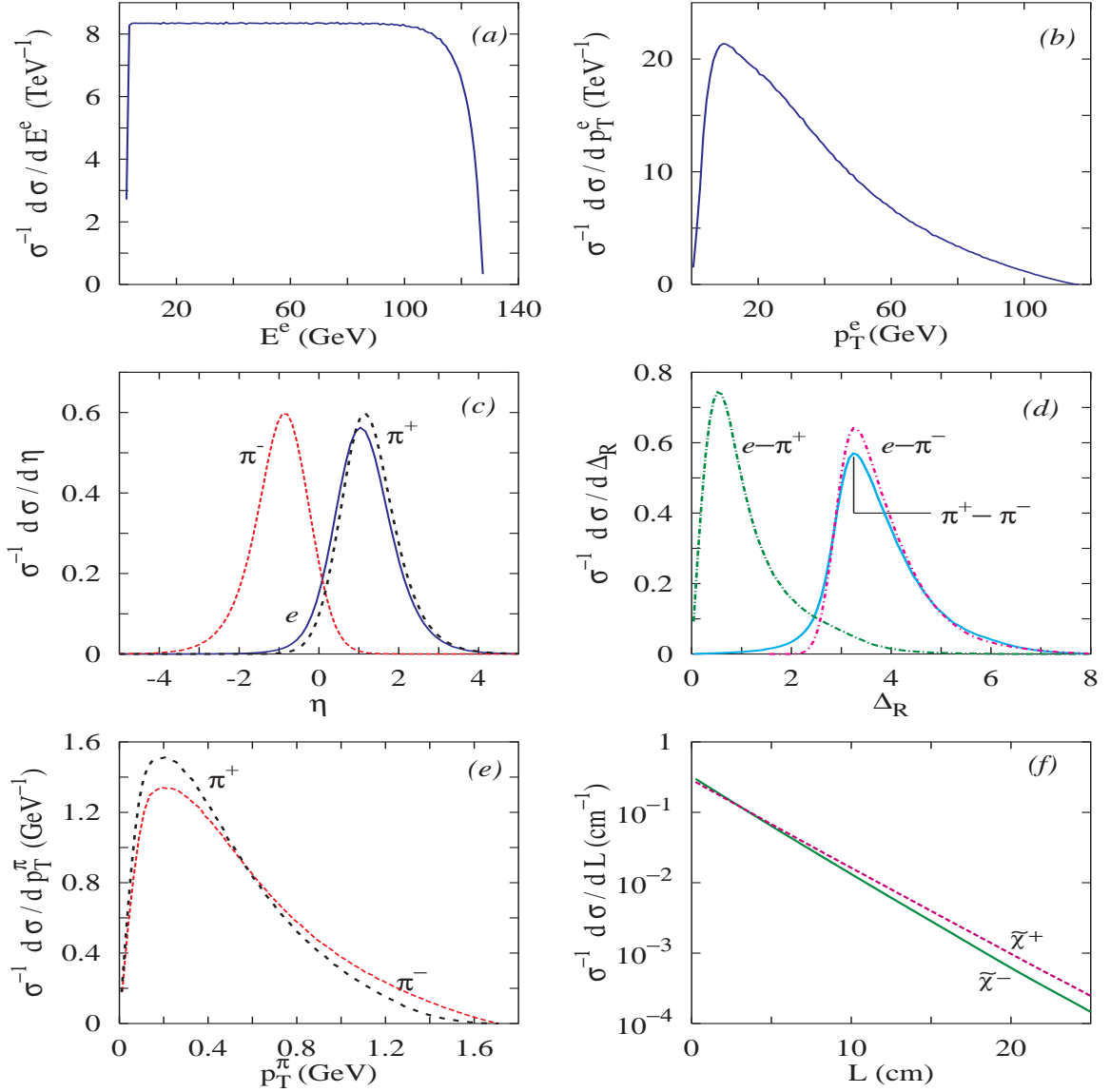}
}
\vspace*{2ex}
\caption{\sf Normalized kinematic distributions of signal events at a
	machine operating with $\sqrt{s}_{ee} = 1 \tev$ and 
	beam polarization choices: $P_L = -1.0, P_{b} = 0.8, P_{e^-} = -0.8$.
	The AMSB parameters are $m_0 = 210~{\rm ~GeV}, m_{3/2} = 37~{\rm~TeV},
        tan\beta = 5, sgn(\mu) > 0$ (point {\bf A}). 
        The different panels correspond to 
	{\em (a)} electron energy; {\em (b)} electron $p_T$; 
        {\em (c)} rapidities;
        {\em (d)} cone separations
        {\em (e)} pion $p_T$s; and
        {\em (f)} chargino decay lengths.
	}
\label{fig:dist}
\end{figure}
Let us concentrate, for the moment, on the parameter space 
point ({\bf A}). Had we a 
monochromatic photon beam, the sneutrino and the chargino would have 
been monochromatic too. And since the electron is a product 
of the two-body decay of the sneutrino into the very same chargino, 
the electron energy would have been strictly bound on both sides
(with the bounds determined by $m_{\tilde \nu}$ and $m_{\Cp1}$), and 
the distribution uniform within. Although this simplistic situation is 
modified to a degree by the spread in the photon spectrum, the main 
feature is still quite discernible (see Fig.~\ref{fig:dist}{\em a}). 
The transverse momentum of the electron (see Fig.~\ref{fig:dist}{\em b}), 
too, shows unmistakable signs of its kinematic origin, with the position 
of the peak being governed, once again, by $m_{\tilde \nu}$ and $m_{\Cp1}$.
With its rapidity distribution mirroring that of the sneutrino, the 
electron is still relatively central, though boosted in the direction of the 
incoming electron beam (Fig.~\ref{fig:dist}{\em c}). 
The same holds for the $\pi^+$ arising from the sneutrino decay. The 
$\pi^-$, on the other hand, is a decay product of the primary chargino
( $ \tilde\chi^-_{1}$) and is hence boosted somewhat in the direction 
of the incoming photon beam (Fig.~\ref{fig:dist}{\em c}). This immediately 
suggests that the $\pi^-$ would be well separated from 
both the electron and the $\pi^+$, a conclusion 
supported by the distribution displayed in Fig.~\ref{fig:dist}{\em d}. 
Once again, this conclusion holds  almost over the entire 
SUSY parameter range under consideration. 
The angular separation between the electron and the  $\pi^+$, on the other 
hand, is a sensitive function of the difference $m_{\tilde \nu} - m_{\Cp1}$
and, for larger mass differences, peaks at larger values.
Of crucial importance are the transverse momenta of the pions, for a minimal
value is necessary for them to be detected. As expected, the pions are 
very soft (see Fig.~\ref{fig:dist}{\em e}), 
with the extent of the $p_T$ (as well 
as energy) distributions determined by $m_{\Cp1} - m_{\N10}$. The very 
same quantity (alongwith the absolute magnitude of the masses) also 
determine the decay length of the charginos. The slightly faster decay 
of the primary chargino in this case (see Fig.~\ref{fig:dist}{\em f})
is but a reflection of its having a lower average momentum.

\begin{table}[htb]
\begin{center}
\footnotesize
\begin{tabular}{|l||c|c|c||c|c|c||}
\hline
(~$m_0~{\rm (GeV)},~ m_{3/2}~{\rm (TeV)}, \tan\beta$~) &
\multicolumn{3}{|c||}{{\bf A.} \hspace*{1em} (210,~37,~5)}   & 
\multicolumn{3}{|c||}{{\bf B.} \hspace*{1em} (450,~55.0,~30)} \\
\hline
(~$m_{\tilde\chi^+_1}, m_{\tilde\chi^0_1}, m_{\tilde\nu}~)~{\rm (GeV)}$
&\multicolumn{3}{|c||}{(~119.64,~119.39,~139.11~)}
  &  \multicolumn{3}{|c||}{(~186.42,~186.23,~390.07)}\\
\hline
& \multicolumn{3}{|c||}{$(P_L, P_b, P_{e^-})$} 
& \multicolumn{3}{|c||}{$(P_L, P_b, P_{e^-})$} \\
\cline{2-7}
& ($-, +, -$) & ($+, -, -$) & ($0, 0, 0$) 
& ($-, +, -$) & ($+, -, -$) & ($0, 0, 0$) \\
\hline
Total $\sigma$ (fb) &  407.1   & 279.3  & 169.3 & 139.3  &  201.9  &  82.3    \\
\hline
$|\eta_e| < 2.5 $   &  386.7   & 250.9  & 157.0 & 136.5  &  197.0  &  80.4 \\
\hline
$p_T^{e} > 20$~GeV  &  265.5   & 144.5  & 100.6 & 136.1  &  196.4  &  80.2 \\
\hline
Both $|\eta_{\pi}| < 2.5 $ 
                    &  248.7   & 125.5  & ~91.6 & 130.7  &  183.2  &  75.9 \\
\hline
Both $p_T^\pi > 200$~MeV     
                    &  179.6   & ~79.5  & ~63.1 & ~30.6  &  ~32.3  &  14.4\\
\hline
$\Delta_{e\pi} > 0.4 $     
                    &  125.5   & ~59.5  & ~45.7 & ~30.3  &  ~32.0  &  14.2\\
\hline
$\Delta_{[\pi_1\pi_2]} > 0.4 $  
                    &  125.5   & ~59.5  & ~45.7 & ~30.3  &  ~32.0  &  14.2\\
\hline
$\mpT > 20$~GeV     &  125.0   & ~59.0  & ~45.4 & ~30.1  &  ~32.0  &  14.2\\
\hline
\multicolumn{1}{|p{1.5in}||}
{Decay length for one chargino $>$ 4.0~cm}
	            &  ~62.5   & ~28.9  & ~21.8 & ~22.1  &  ~22.2  &   9.68\\
\hline
\multicolumn{1}{|p{1.5in}||}
{Decay length for both chargino $>$ 4.0~cm}
	            &  ~10.7   & ~4.6  & ~3.48 & ~6.9  &  ~6.37  &   2.70  \\
\hline
\end{tabular}
\end{center}

\caption{\sf Illustrating the effects of various cuts on the signal  
cross-sections for two specimen points in the parameter space and  
for $\sqrt{s}_{ee}= 1$ TeV. In either case, $\mu > 0$, and 
the cross sections are shown both for  unpolarized beams 
as well as for two particular choices for beam polarizations.
Whenever nonzero, $|P_L|$ = 1 and $|P_b| = |P_{e^-}| = 0.8$.}
\label{table:cuts}
\end{table}
Having discussed the kinematics, it is now relatively easy to 
anticipate the effects of the cuts that we have chosen to impose. 
As we shall shortly see, the efficiency has a nontrivial dependence 
on the superparticle masses. While, in an actual experimental context,
it might be more prudent to employ flexible cuts, we shall, nevertheless, 
continue to work with those that we have already described. 
As Table~\ref{table:cuts} shows, the requirement on the 
minimum electron transverse momentum serves to reject quite a few 
signal events for a small $m_{\tilde \nu}$ and/or a small difference 
$m_{\tilde \nu} - m_{\Cp1}$. Were both these to be large (point {\bf B}),
the effect is marginal at best. A similar statement may also be made for 
the angular separation between the electron and the $\pi^+$. 
On the other hand, a small mass difference 
between the chargino and neutralino renders the requirement on the 
pion transverse momentum a crucial one (a larger fraction of the signal
is lost for point {\bf B} than for point {\bf A}). And, as expected, the 
rest of the cuts imposed have only marginal effect on the signal. Finally, 
a remark on the chargino track lengths. As shown in Table.~\ref{table:cuts}, 
if we insist that both charginos leave a substantial track, we would be 
discarding the majority of events. While this may still be fine for 
relatively lighter superparticles, when the cross sections are large, such a 
requirement would severely limit the reach of the experiment. In view of our 
demonstration that our signal is relatively background-free even in the 
absence of macroscopic tracks, it is perhaps unnecessary to impose such a 
requirement, at least in the discovery stage. Of course, if a signal were to 
appear, one could always go back and look for the confirmatory tracks. This 
is the standpoint we shall adopt in the remainder of the article.

\subsection{Signal strength and the parameter space}

Until now, we have restricted ourselves to an examination of the 
signal profile for two specific points 
in the parameter space. While the gross features can be inferred from 
the discussion in the previous section, it is of interest to determine 
magnitude of the total signal strength as a function of the parameters
involved. We aim to do this now. 
Clearly, it is not possible to depict the dependence on each of the 
parameters, and hence we shall restrict ourselves to two discrete values 
of $\tan \beta$ while allowing $m_0$ and $m_{3/2}$ to vary freely 
modulo the experimental constraints. As for the beam polarization, we 
make a particular choice, namely $P_L = +1, P_{b} = P_{e^-} = -0.8$. 

\begin{figure}[!ht]
\vspace*{8ex}
\centerline{
\epsfxsize=9.5cm\epsfysize=9.5cm
                    \epsfbox{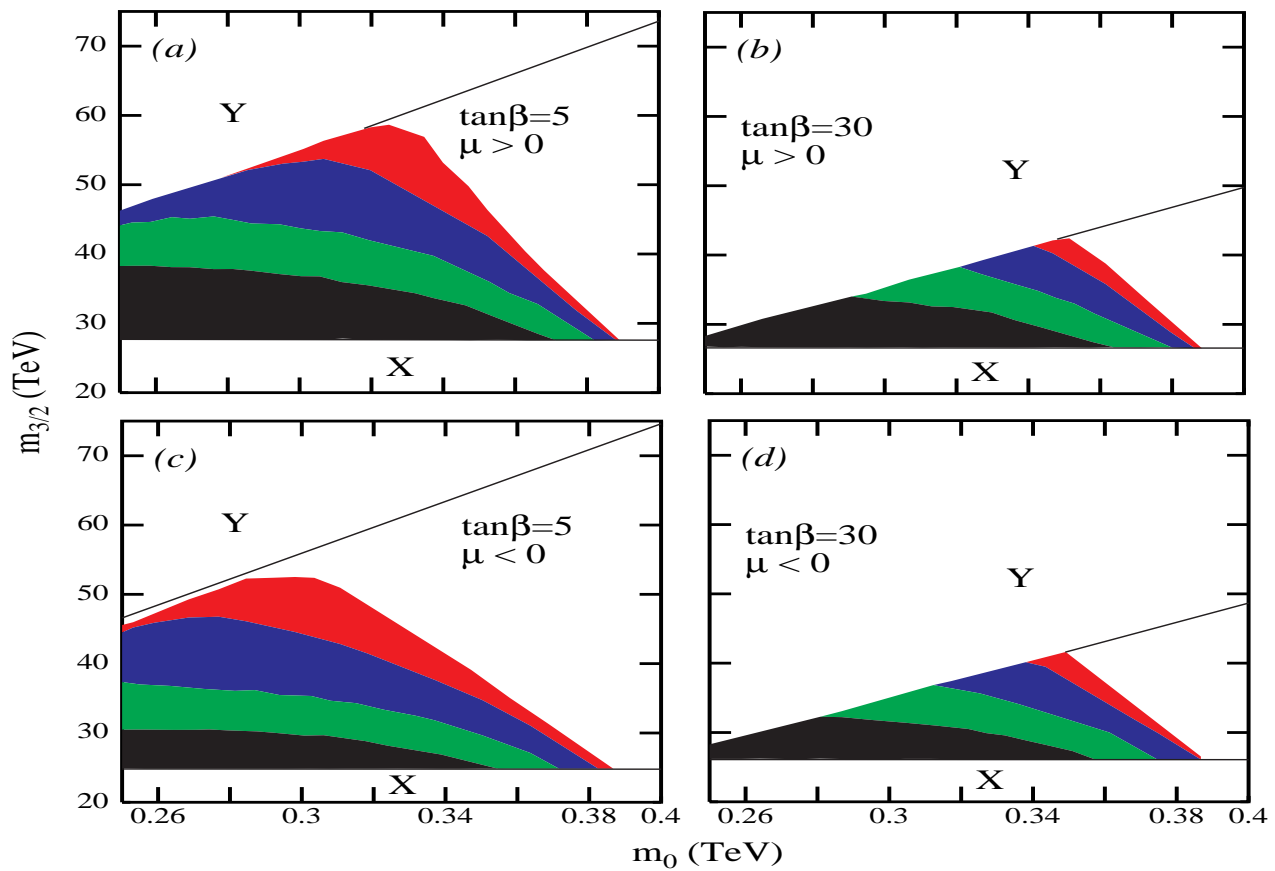}
}
\caption{\sf Scatter plots for the signal cross-section (fb) in the
         $m_0-m_{3/2}$ plane for a machine operating 
         at $\sqrt{s}_{ee} = 500\gev$ and $P_L = +1, P_{b} = P_{e^-} = -0.8$.
	The values of $tan\beta$ and $ sgn(\mu)$
         are as indicated. The regions marked by 
        $X$ are ruled out by the experimental lower limit on the chargino 
	mass, while those marked by $Y$ are ruled out by the requirement 
	of $\N10$ being the LSP.
         In each panel, the top three shaded 
	 regions correspond to cross section ranges of 
         $[(0.1-5),~(5-50),~(50-150)]$. The lowermost region 
         corresponds to $(150-470)$ in ${\em (a)}$,
         $(150-390)$ in ${\em (b)}$, $(150-335)$ in ${\em (c)}$
	 and $(150-350)$ in ${\em (d)}$ respectively.}
\label{fig-500}
\end{figure}

In Fig.\ref{fig-500}, we display our results for a machine 
operating at $\sqrt{s}_{ee} = 500 \gev$, in the form of scatter 
plots for the cross section, after imposing the cuts,
in the plane spanned by $m_0$ and $m_{3/2}$.
The panels on the left(right) correspond to low (high) $\tan \beta$
(5 and 30). For each choice, we also depict the dependence on the sign 
of the $\mu$-term. In each of the individual graphs, the region marked 
by $X$ corresponds to a chargino mass of less than 86 GeV, thereby 
falling foul of the ALEPH bound~\cite{aleph}. The region $Y$, on the other 
hand, would correspond to the $\tilde \tau_1$ being the LSP, a possibility
not allowed phenomenologically if $R$-parity is to be conserved. 
We, thus,
need to consider ourselves only with the wedge-shaped region enclosed by 
the two straight lines. Note that 
the extent of the region $Y$ is much more sensitive to the value of 
$\tan \beta$ than is the case for the region $X$. This is easy to understand 
when one considers the fact that a large $\tan\beta $ results in making 
the $\tau$-Yukawa coupling stronger, which, in turn, drives down
the mass of the lighter $\tilde\tau$, thereby  rendering it the LSP.

It is obvious that
the total cross section would fall monotonically with each of the 
two mass parameters, and this is amply reflected in the figure. Note 
that the variation with $m_{3/2}$ is much sharper for large $m_0$. This,
again, is easy to understand as one is progressively squeezing the 
available phase space for the production process. The 
variations with  $\tan\beta$ and the sign of $\mu$ are of a subtler 
origin, namely
the dependence of $\Delta M$ (the mass-difference  between the lighter 
chargino and the lightest neutralino) on these two parameters. 
It can be easily ascertained that, for positive $\mu$, the 
mass-difference $\Delta M$ decreases with $\tan\beta$. This, in turn,
implies that the pions resulting from the decay of the chargino 
have a higher average transverse momentum for lower $\tan\beta$, thereby 
making them easier to detect. For negative $\mu$, the effect is opposite. 

\begin{figure}[!ht]
\vspace*{8ex}
\centerline{
\epsfxsize=9.5cm\epsfysize=9.5cm
                    \epsfbox{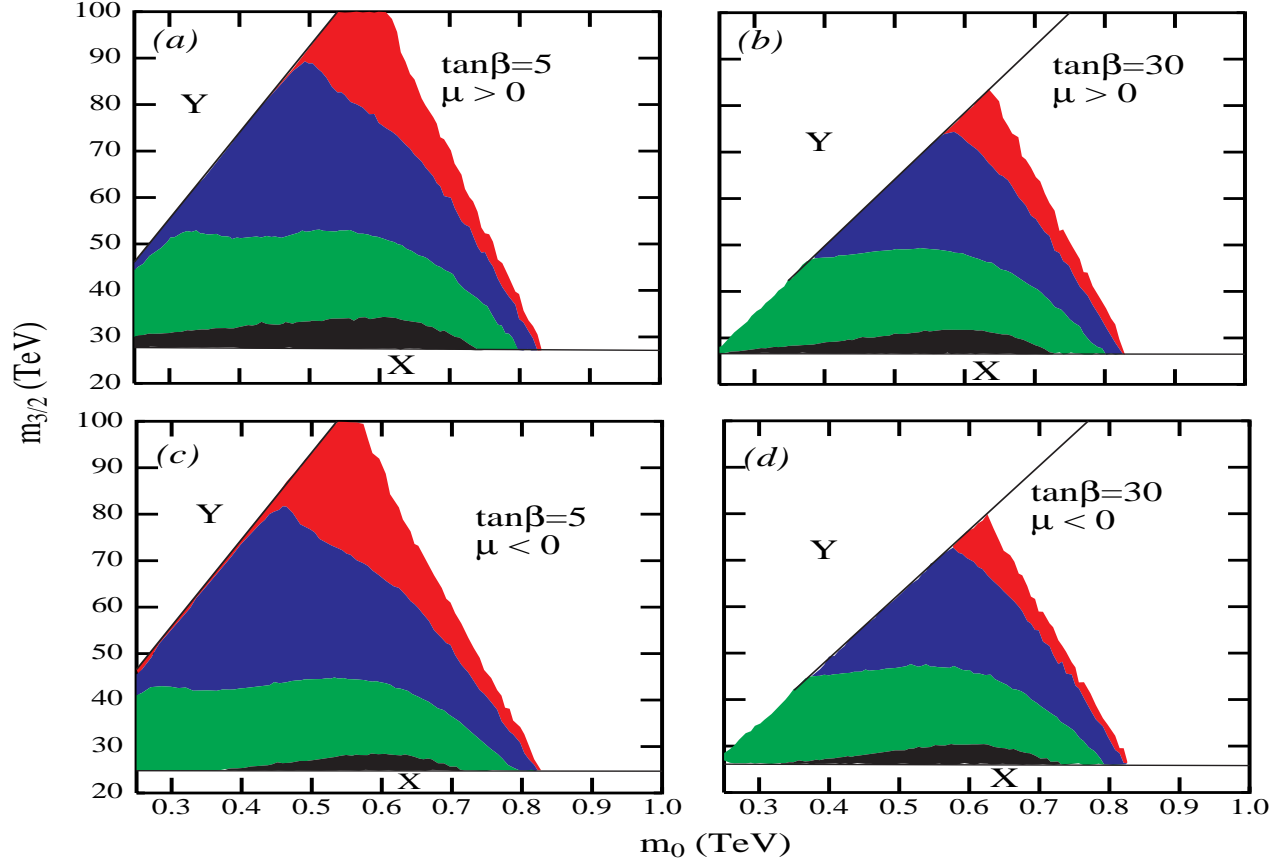}
}
\caption{\sf As in Fig.\protect\ref{fig-500}, but for 
	$\sqrt{s}_{ee} = 1 \tev$ instead. In each panel, the top three shaded 
	 regions correspond to cross section ranges of 
         $[(0.1-5),~(5-50),~(50-150)]$. The lowermost region 
         corresponds to $(150-215)$ in ${\bf (a)}$,
         $(150-205)$ in ${\bf (b)}$, $(150-185)$ in ${\bf (c)}$
	 and $(150-195)$ in ${\bf (d)}$ respectively.}
\label{fig-1tev}
\end{figure}

It is clear from Fig.\ref{fig-500} that, for the low $tan\beta $
case, an experiment such as this can easily explore 
$m_{3/2}$ as high as $60 (50)$~TeV 
for negative (positive) $\mu$.
For $tan\beta = 30$, on the other hand, the maximal reach in $m_{3/2}$ 
is approximately 40 TeV irrespective of $sgn(\mu)$. 
Similarly, the reach in $m_0$ has little dependence on either of 
$tan\beta$ and $sgn(\mu)$.
Finally, it must be borne in mind that Fig.\ref{fig-500} has
been drawn keeping in mind a moderate luminosity ($\lsim 100 \fb^{-1}$).
A significantly larger integrated  luminosity would, of course, allow 
one to explore beyond the topmost shaded band. 

In Fig.\ref{fig-1tev}, we show a similar plot for a machine opearting at
$\sqrt{s}_{ee} = 1 \tev$ instead. The features are very similar, with the 
obvious enhancement in the reach.

\section{Parameter Determination}

Once the existence of a new particle has been confirmed, it is 
almost contingent upon the experimentalist to determine its mass, spin 
etc. and, hopefully, glean further information as to the possible 
values of the parameters within a particular theoretical framework. 
We shall now investigate the possibility for such studies at an $e \gamma$ 
collider. 

Positing that the particles produced were the sneutrino and the chargino 
and that the sneutrino subsequently decayed into a similar chargino and 
an electron, it is easy to see that the energy of the decay electron is 
strictly confined~\cite{choudhury} within the interval 
\begin{equation}
{m_{\tilde \nu}^2-m_{\tilde\chi_1^+}^2
\over
2\left(
E_{\tilde \nu}^{\rm max} + k_{\tilde \nu}^{\rm max}
\right)}
\leq E^e \leq
{m_{\tilde \nu}^2-m_{\tilde\chi_1^+}^2
\over
2\left(
E_{\tilde \nu}^{\rm max} - k_{\tilde \nu}^{\rm max}
\right)}
\ ,
\label{e_elec}
\end{equation}
where $E_{\tilde \nu}^{\rm max}$ is the maximum possible energy that 
the intermediate sneutrino may have carried, viz.
\begin{eqnarray}
E_{\tilde \nu}^{\rm max}
=
{1 \over 4y_{\rm max}\sqrt{s}}
&\Biggl[&
(1+y_{\rm max}) \left(y_{\rm max}s+m_{\tilde \nu}^2-m_{\tilde\chi_1^+}^2\right)
\nonumber\\
&+&
(1-y_{\rm max})
        \sqrt{ \left(y_{\rm max}s+m_{\tilde \nu}^2-m_{\tilde\chi_1^+}^2\right)^2
                - 4y_{\rm max}sm_{\tilde \nu}^2 }
\quad\Biggr] \ ,
\label{e_sneu}
\end{eqnarray}
and $k_{\tilde \nu}^{\rm max}$ is the corresponding momentum.

\begin{figure}[!ht]
\vspace*{-16ex}
\centerline{
\epsfxsize=8cm\epsfysize=8.0cm
                    \epsfbox{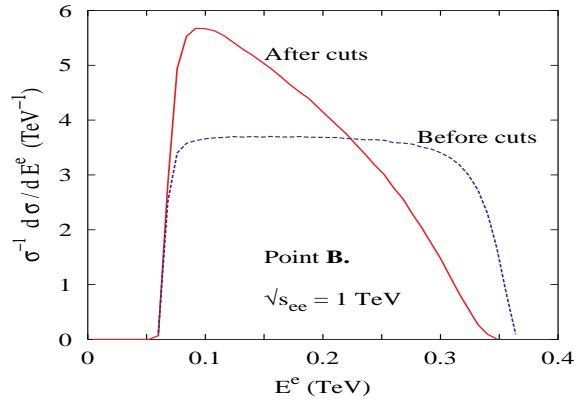}
}
\caption[]{\sf The normalized electron energy distribution for parameter 
	       point {\bf B} (see text). The dashed (solid) lines 
	       depict the cross section before (after) 
	       {\em all} the selection criteria are imposed.}
\label{fig:obs_e}
\end{figure}

As $\sqrt{s}$ and $y_{\rm max}$ are both known, it follows that an accurate
measurement of $E^e_{\rm max}$ and  $E^e_{\rm min}$ would allow us to 
determine both the masses. A very precise measurement of the endpoints is
unlikely, though. The reasons are manifold. Apart from the issue of energy 
resolution, one has to take into account that the sharp edges, as displayed
in Fig.~\ref{fig:dist}{\em a}, do {\em not} remain as sharp once 
{\em all} the cuts are imposed. Rather, there is a more gradual tapering 
off at the higher end, as displayed in Fig.~\ref{fig:obs_e}. This, obviously,
could result in an underestimation of $E^e_{\rm max}$, and, from a random 
survey in the parameter space, we find that this effect is typically 
$\lsim 20 \gev$. To account for this, we consider a conservative
error contribution of $25 \gev$ in the measurement of $E^e_{\rm max}$. 

The determination of $E^e_{\rm min}$, on the other hand, does not suffer 
from this problem, and, hence, we only consider an error of $5 \gev$ 
to account for the finite resolution. There is a caveat, though. Looking at 
Fig.~\ref{fig:dist}{\em a} 
(or, equivalently, eqns.~\ref{e_elec}\&\ref{e_sneu}), 
one immediately sees that the true $E^e_{\rm min}$ may very well be less than 
$20 \gev$, and hence in contradiction with our requirement on the minimal 
transverse momentum for the electron. Clearly, for such points in the 
parameter space, $\delta E^e_{\rm min} \gsim 20 \gev$. Of course, for a 
given collider configuration, 
it is easy to determine the part of the parameter space that is beset with 
this particular problem. In the rest of this section, we rather concentrate 
on a point that lies {\em outside} this region, namely the point {\bf B} 
discussed earlier.

\begin{figure}[!ht]
\vspace*{-20ex}
\centerline{
\epsfxsize=8cm\epsfysize=10.0cm
                    \epsfbox{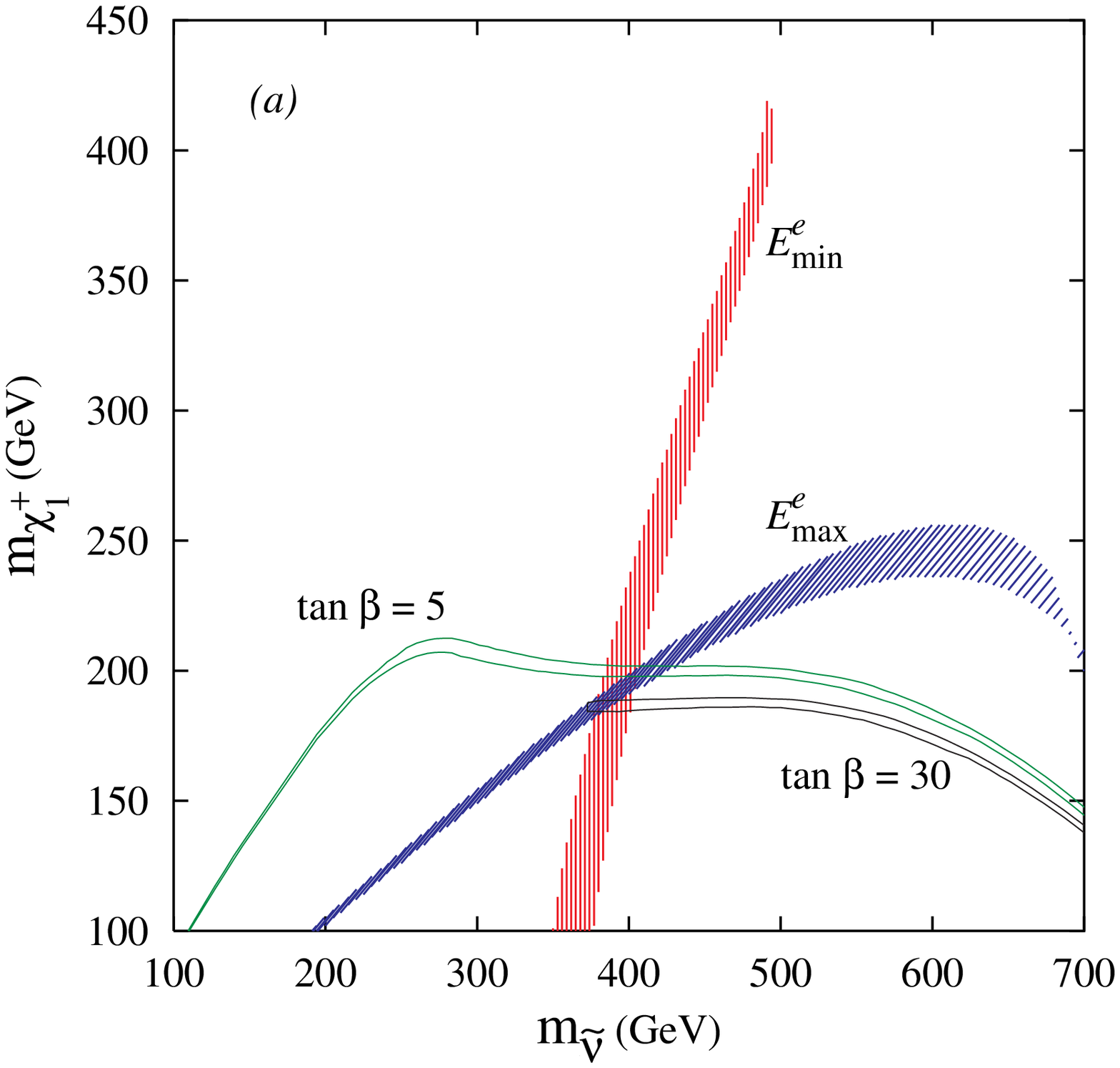}
\epsfxsize=8cm\epsfysize=10.0cm
                    \epsfbox{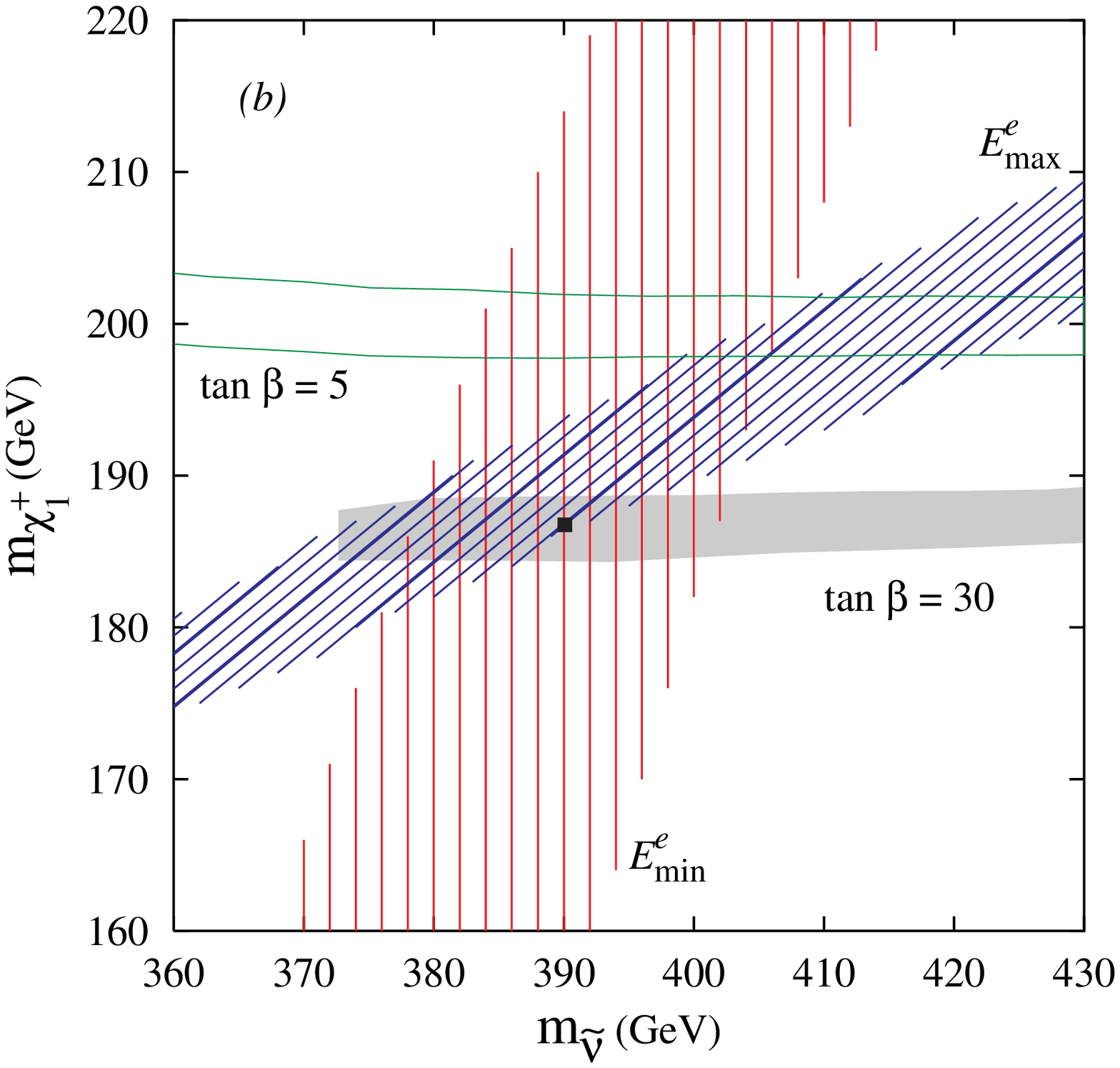}
}
\caption[]{\sf {\em (a)} The determination of the sneutrino and chargino 
	masses from a measurement of the endpoints of the electron energy
	spectrum (eqn.\protect\ref{e_elec}). The width of the bands correspond
	to the error bars described in the text. For points within 
	the two horizontal bands, the resultant cross section would agree 
	with the measured one to within $1 \sigma$. 
	{\em (b)} The overlap region has been shown on an expanded scale. 
	The dark square denotes the reference point in the parameter 
	space (pt. {\bf B} in the text).
	}
\label{fig:mass_det}
\end{figure}

In Fig.~\ref{fig:mass_det}{\em a}, we display the two bands obtained from 
the measurement of the two electron energy endpoints corresponding to 
point {\bf B}. The measurement errors are exactly as described above, and 
the curious shape of the band is but a consequence of the nonlinear nature 
of eqn.(\ref{e_elec}). What is interesting is that the two measurements 
lead to very different constraints in the parameter space, thereby 
facilitating a relatively good measurement of the two masses. The error 
on the sneutrino mass, thus determined, is roughly 12 GeV, while that 
on the chargino mass is roughly 6 GeV. Moreover, the errors are quite 
correlated (see Fig.~\ref{fig:mass_det}{\em b}, which displays the 
region of interest on an expanded scale) and hence the 
combined $1 \sigma$ ellipse would lie well within the overlap region.
Note that all this 
information has been gleaned from a single process and {\em without}
an energy scanning.

Having determined the masses, it is now of interest to measure the 
remaining parameter, namely $\tan \beta$. Clearly, kinematical distributions
are essentially independent of this quantity and one should rather consider 
cross section measurement, or, in other words, a counting experiment. 
For this purpose, we choose to work with a moderate choice of 
luminosity, viz. ${\cal L} = 100 \fb^{-1}$. 
The cross sections are converted to event numbers through the relation
\[
	n = \sigma \cdot \epsilon \cdot {\cal L}
\]
where the overall detection efficiency $\epsilon$ is but the product of the 
efficiency for pion detection (assumed to be 50\% each) and those for 
electron detection (95\%). 
The corresponding error in the cross section 
measurement (after imposing the cuts, naturally) is easily determined 
on application of Poisson (or Gaussian) statistics. 
Armed with this, and for a given value of 
$\tan \beta$, one could easily determine the part of the 
$m_{\tilde \nu}$--$m_{\tilde \chi_1^+}$ space that would be consistent 
with the measured cross section. In Figs.~\ref{fig:mass_det}, we display 
these constraints for two particular values of the ratio $\tan \beta$. One 
might wonder at the abrupt end for the band corresponding to 
$\tan \beta = 30$. This, however, is but a consequence of the aforementioned 
constraints on the mass of $\tilde \tau_{1}$. Note that, while some 
resolution is possible, such experiments are not overly sensitive to this 
parameter. It is possible that significant improvement would occur once other 
production processes are considered, but that is beyond the scope of the 
present work. 

\section{Conclusions}

To summarise, we discuss the feasibility of using an electron-photon collider 
to investigate minimal models wherein supersymmetry breaking is mediated 
to the visble sector through the super-Weyl anomaly. A very 
striking feature of such models, 
including the minimal model we have studied here, is that the lightest 
chargino $\C1pm$ and the lightest neutralino $\N10$ are nearly degenerate 
and predominantly Winos. This leads to a long-lived $\C1pm$ which then decays 
into $\N10 + \pi^\pm$ (soft) resulting in a heavily ionizing charged track 
and a soft $\pi^\pm$ whose impact parameter may be measurable. While 
signals for such scenarios have been studied in the context of other colliders,
this very feature often restricts the detectability of the model.

We demonstrated though, that the associated production of the lighter 
chargino and the sneutrino at an $e\gamma$ collider 
could provide a very clean signature for such a scenario. 
The signal event 
consists of an energetic electron (emanating from the decay of the sneutrino) 
which serves as the trigger, two macroscopic charged tracks in the vertex 
detector and/or two soft charged pions and, of course, a large missing 
transverse momentum. 
The possible SM backgrounds are calculated and shown to be 
negligibly small once suitable selection criteria are employed. 
Consequently, even 
with an integrated luminosity of only $100 \fb^{-1}$, 
one could see as many as 1000 (background-free) signal events 
over a very large region of the allowed parameter space. 

Additional advantages of the mode we advocate are encapsulated 
in the kinematic distribution of the signal events. The presence of 
distinct energy endpoints for the electron allows one to 
determine the masses of both the chargino and the sneutrino to a fair degree 
of accuracy. For a very large part of the parameter space, this is true
even {\em without} an energy scan. The latter technique, if employed, 
can only improve the measurements. In addition, a measurement of the 
cross section can be used to constrain the possible range for
$\tan \beta$. 

Once the signal has been established and all possible information gleaned, 
the same collider can easily be used to obtain confirmatory checks. 
The simplest example is the associated production of the left-selectron 
with the lightest neutralino. A more non-trivial example (and requiring 
a higher energy) is the associated production of the right-selectron 
with the second-lightest neutralino. Using techniques similar to those 
discussed here, one could also measure the masses of these two particles. 
Since AMSB models have very definite predictions for the ratios of the 
masses of two lightest neutralinos as well as for the splitting of 
slepton masses, such tests are likely to be crucial in 
establishing the mechanism of supersymmetry breaking as well 
as the parameters of the theory. We hope to come back to this issue 
in the near future. 

{\bf Acknowledgments}
DKG would like thank the Department of Physics, National Taiwan University,
Taipei, where the initial part of this work was done. 
SR thanks
Aseshkrishna Datta and Biswarup Mukhopadhyaya for very helpful
discussions. DC would like to thank the 
Dept. of Science and Technology, India for financial assistance under 
the Swarnajayanti Fellowship grant.

\newpage

\def\pr#1,#2 #3 { {Phys.~Rev.}        ~{\bf #1},  #2 (19#3)}
\def\prd#1,#2 #3{ { Phys.~Rev.}       ~{D \bf #1}, #2 (19#3)}
\def\pprd#1,#2 #3{ { Phys.~Rev.}      ~{D \bf #1}, #2 (20#3)}
\def\prl#1,#2 #3{ { Phys.~Rev.~Lett.}  ~{\bf #1},  #2 (19#3)}
\def\pprl#1,#2 #3{ {Phys. Rev. Lett.}   {\bf #1},  #2 (20#3)}
\def\plb#1,#2 #3{ { Phys.~Lett.}       ~{B \bf #1}, #2 (19#3)}
\def\pplb#1,#2 #3{ {Phys. Lett.}        {B \bf #1}, #2 (20#3)}
\def\npb#1,#2 #3{ { Nucl.~Phys.}       ~{\bf B#1}, #2 (19#3)}
\def\pnpb#1,#2 #3{ {Nucl. Phys.}        {\bf B#1}, #2 (20#3)}
\def\prp#1,#2 #3{ { Phys.~Rep.}       ~{\bf #1},  #2 (19#3) }
\def\zpc#1,#2 #3{ { Z.~Phys.}          ~{\bf C#1}, #2 (19#3)}
\def\epj#1,#2 #3{ { Eur.~Phys.~J.}     ~{\bf C#1}, #2 (19#3)}
\def\mpl#1,#2 #3{ { Mod.~Phys.~Lett.}  ~{\bf A#1}, #2 (19#3)}
\def\ijmp#1,#2 #3{{ Int.~J.~Mod.~Phys.}~{\bf A#1}, #2 (19#3)}
\def\ptp#1,#2 #3{ { Prog.~Theor.~Phys.}~{\bf #1},  #2 (19#3)}
\def\jhep#1, #2 #3{ {J. High Energy Phys.} {\bf #1}, #2 (19#3)}
\def\pjhep#1, #2 #3{ {J. High Energy Phys.} {\bf #1}, #2 (20#3)}

\end{document}